\documentclass[letterpaper,compsoc,twoside]{IEEEtran}
\usepackage{fixltx2e} 
\usepackage{cmap} 
\usepackage{ifthen}
\usepackage[T1]{fontenc}
\usepackage[utf8]{inputenc}

\usepackage[font={small,it},labelfont=bf]{caption}
\usepackage{float}

\pdfoutput=1
\usepackage{scipy}
\makeatletter
\def\PY@reset{\let\PY@it=\relax \let\PY@bf=\relax%
    \let\PY@ul=\relax \let\PY@tc=\relax%
    \let\PY@bc=\relax \let\PY@ff=\relax}
\def\PY@tok#1{\csname PY@tok@#1\endcsname}
\def\PY@toks#1+{\ifx\relax#1\empty\else%
    \PY@tok{#1}\expandafter\PY@toks\fi}
\def\PY@do#1{\PY@bc{\PY@tc{\PY@ul{%
    \PY@it{\PY@bf{\PY@ff{#1}}}}}}}
\def\PY#1#2{\PY@reset\PY@toks#1+\relax+\PY@do{#2}}

\expandafter\def\csname PY@tok@gd\endcsname{\def\PY@tc##1{\textcolor[rgb]{0.63,0.00,0.00}{##1}}}
\expandafter\def\csname PY@tok@gu\endcsname{\let\PY@bf=\textbf\def\PY@tc##1{\textcolor[rgb]{0.50,0.00,0.50}{##1}}}
\expandafter\def\csname PY@tok@gt\endcsname{\def\PY@tc##1{\textcolor[rgb]{0.00,0.27,0.87}{##1}}}
\expandafter\def\csname PY@tok@gs\endcsname{\let\PY@bf=\textbf}
\expandafter\def\csname PY@tok@gr\endcsname{\def\PY@tc##1{\textcolor[rgb]{1.00,0.00,0.00}{##1}}}
\expandafter\def\csname PY@tok@cm\endcsname{\let\PY@it=\textit\def\PY@tc##1{\textcolor[rgb]{0.25,0.50,0.56}{##1}}}
\expandafter\def\csname PY@tok@vg\endcsname{\def\PY@tc##1{\textcolor[rgb]{0.73,0.38,0.84}{##1}}}
\expandafter\def\csname PY@tok@vi\endcsname{\def\PY@tc##1{\textcolor[rgb]{0.73,0.38,0.84}{##1}}}
\expandafter\def\csname PY@tok@mh\endcsname{\def\PY@tc##1{\textcolor[rgb]{0.13,0.50,0.31}{##1}}}
\expandafter\def\csname PY@tok@cs\endcsname{\def\PY@tc##1{\textcolor[rgb]{0.25,0.50,0.56}{##1}}\def\PY@bc##1{\setlength{\fboxsep}{0pt}\colorbox[rgb]{1.00,0.94,0.94}{\strut ##1}}}
\expandafter\def\csname PY@tok@ge\endcsname{\let\PY@it=\textit}
\expandafter\def\csname PY@tok@vc\endcsname{\def\PY@tc##1{\textcolor[rgb]{0.73,0.38,0.84}{##1}}}
\expandafter\def\csname PY@tok@il\endcsname{\def\PY@tc##1{\textcolor[rgb]{0.13,0.50,0.31}{##1}}}
\expandafter\def\csname PY@tok@go\endcsname{\def\PY@tc##1{\textcolor[rgb]{0.20,0.20,0.20}{##1}}}
\expandafter\def\csname PY@tok@cp\endcsname{\def\PY@tc##1{\textcolor[rgb]{0.00,0.44,0.13}{##1}}}
\expandafter\def\csname PY@tok@gi\endcsname{\def\PY@tc##1{\textcolor[rgb]{0.00,0.63,0.00}{##1}}}
\expandafter\def\csname PY@tok@gh\endcsname{\let\PY@bf=\textbf\def\PY@tc##1{\textcolor[rgb]{0.00,0.00,0.50}{##1}}}
\expandafter\def\csname PY@tok@ni\endcsname{\let\PY@bf=\textbf\def\PY@tc##1{\textcolor[rgb]{0.84,0.33,0.22}{##1}}}
\expandafter\def\csname PY@tok@nl\endcsname{\let\PY@bf=\textbf\def\PY@tc##1{\textcolor[rgb]{0.00,0.13,0.44}{##1}}}
\expandafter\def\csname PY@tok@nn\endcsname{\let\PY@bf=\textbf\def\PY@tc##1{\textcolor[rgb]{0.05,0.52,0.71}{##1}}}
\expandafter\def\csname PY@tok@no\endcsname{\def\PY@tc##1{\textcolor[rgb]{0.38,0.68,0.84}{##1}}}
\expandafter\def\csname PY@tok@na\endcsname{\def\PY@tc##1{\textcolor[rgb]{0.25,0.44,0.63}{##1}}}
\expandafter\def\csname PY@tok@nb\endcsname{\def\PY@tc##1{\textcolor[rgb]{0.00,0.44,0.13}{##1}}}
\expandafter\def\csname PY@tok@nc\endcsname{\let\PY@bf=\textbf\def\PY@tc##1{\textcolor[rgb]{0.05,0.52,0.71}{##1}}}
\expandafter\def\csname PY@tok@nd\endcsname{\let\PY@bf=\textbf\def\PY@tc##1{\textcolor[rgb]{0.33,0.33,0.33}{##1}}}
\expandafter\def\csname PY@tok@ne\endcsname{\def\PY@tc##1{\textcolor[rgb]{0.00,0.44,0.13}{##1}}}
\expandafter\def\csname PY@tok@nf\endcsname{\def\PY@tc##1{\textcolor[rgb]{0.02,0.16,0.49}{##1}}}
\expandafter\def\csname PY@tok@si\endcsname{\let\PY@it=\textit\def\PY@tc##1{\textcolor[rgb]{0.44,0.63,0.82}{##1}}}
\expandafter\def\csname PY@tok@s2\endcsname{\def\PY@tc##1{\textcolor[rgb]{0.25,0.44,0.63}{##1}}}
\expandafter\def\csname PY@tok@nt\endcsname{\let\PY@bf=\textbf\def\PY@tc##1{\textcolor[rgb]{0.02,0.16,0.45}{##1}}}
\expandafter\def\csname PY@tok@nv\endcsname{\def\PY@tc##1{\textcolor[rgb]{0.73,0.38,0.84}{##1}}}
\expandafter\def\csname PY@tok@s1\endcsname{\def\PY@tc##1{\textcolor[rgb]{0.25,0.44,0.63}{##1}}}
\expandafter\def\csname PY@tok@ch\endcsname{\let\PY@it=\textit\def\PY@tc##1{\textcolor[rgb]{0.25,0.50,0.56}{##1}}}
\expandafter\def\csname PY@tok@m\endcsname{\def\PY@tc##1{\textcolor[rgb]{0.13,0.50,0.31}{##1}}}
\expandafter\def\csname PY@tok@gp\endcsname{\let\PY@bf=\textbf\def\PY@tc##1{\textcolor[rgb]{0.78,0.36,0.04}{##1}}}
\expandafter\def\csname PY@tok@sh\endcsname{\def\PY@tc##1{\textcolor[rgb]{0.25,0.44,0.63}{##1}}}
\expandafter\def\csname PY@tok@ow\endcsname{\let\PY@bf=\textbf\def\PY@tc##1{\textcolor[rgb]{0.00,0.44,0.13}{##1}}}
\expandafter\def\csname PY@tok@sx\endcsname{\def\PY@tc##1{\textcolor[rgb]{0.78,0.36,0.04}{##1}}}
\expandafter\def\csname PY@tok@bp\endcsname{\def\PY@tc##1{\textcolor[rgb]{0.00,0.44,0.13}{##1}}}
\expandafter\def\csname PY@tok@c1\endcsname{\let\PY@it=\textit\def\PY@tc##1{\textcolor[rgb]{0.25,0.50,0.56}{##1}}}
\expandafter\def\csname PY@tok@o\endcsname{\def\PY@tc##1{\textcolor[rgb]{0.40,0.40,0.40}{##1}}}
\expandafter\def\csname PY@tok@kc\endcsname{\let\PY@bf=\textbf\def\PY@tc##1{\textcolor[rgb]{0.00,0.44,0.13}{##1}}}
\expandafter\def\csname PY@tok@c\endcsname{\let\PY@it=\textit\def\PY@tc##1{\textcolor[rgb]{0.25,0.50,0.56}{##1}}}
\expandafter\def\csname PY@tok@mf\endcsname{\def\PY@tc##1{\textcolor[rgb]{0.13,0.50,0.31}{##1}}}
\expandafter\def\csname PY@tok@err\endcsname{\def\PY@bc##1{\setlength{\fboxsep}{0pt}\fcolorbox[rgb]{1.00,0.00,0.00}{1,1,1}{\strut ##1}}}
\expandafter\def\csname PY@tok@mb\endcsname{\def\PY@tc##1{\textcolor[rgb]{0.13,0.50,0.31}{##1}}}
\expandafter\def\csname PY@tok@ss\endcsname{\def\PY@tc##1{\textcolor[rgb]{0.32,0.47,0.09}{##1}}}
\expandafter\def\csname PY@tok@sr\endcsname{\def\PY@tc##1{\textcolor[rgb]{0.14,0.33,0.53}{##1}}}
\expandafter\def\csname PY@tok@mo\endcsname{\def\PY@tc##1{\textcolor[rgb]{0.13,0.50,0.31}{##1}}}
\expandafter\def\csname PY@tok@kd\endcsname{\let\PY@bf=\textbf\def\PY@tc##1{\textcolor[rgb]{0.00,0.44,0.13}{##1}}}
\expandafter\def\csname PY@tok@mi\endcsname{\def\PY@tc##1{\textcolor[rgb]{0.13,0.50,0.31}{##1}}}
\expandafter\def\csname PY@tok@kn\endcsname{\let\PY@bf=\textbf\def\PY@tc##1{\textcolor[rgb]{0.00,0.44,0.13}{##1}}}
\expandafter\def\csname PY@tok@cpf\endcsname{\let\PY@it=\textit\def\PY@tc##1{\textcolor[rgb]{0.25,0.50,0.56}{##1}}}
\expandafter\def\csname PY@tok@kr\endcsname{\let\PY@bf=\textbf\def\PY@tc##1{\textcolor[rgb]{0.00,0.44,0.13}{##1}}}
\expandafter\def\csname PY@tok@s\endcsname{\def\PY@tc##1{\textcolor[rgb]{0.25,0.44,0.63}{##1}}}
\expandafter\def\csname PY@tok@kp\endcsname{\def\PY@tc##1{\textcolor[rgb]{0.00,0.44,0.13}{##1}}}
\expandafter\def\csname PY@tok@w\endcsname{\def\PY@tc##1{\textcolor[rgb]{0.73,0.73,0.73}{##1}}}
\expandafter\def\csname PY@tok@kt\endcsname{\def\PY@tc##1{\textcolor[rgb]{0.56,0.13,0.00}{##1}}}
\expandafter\def\csname PY@tok@sc\endcsname{\def\PY@tc##1{\textcolor[rgb]{0.25,0.44,0.63}{##1}}}
\expandafter\def\csname PY@tok@sb\endcsname{\def\PY@tc##1{\textcolor[rgb]{0.25,0.44,0.63}{##1}}}
\expandafter\def\csname PY@tok@k\endcsname{\let\PY@bf=\textbf\def\PY@tc##1{\textcolor[rgb]{0.00,0.44,0.13}{##1}}}
\expandafter\def\csname PY@tok@se\endcsname{\let\PY@bf=\textbf\def\PY@tc##1{\textcolor[rgb]{0.25,0.44,0.63}{##1}}}
\expandafter\def\csname PY@tok@sd\endcsname{\let\PY@it=\textit\def\PY@tc##1{\textcolor[rgb]{0.25,0.44,0.63}{##1}}}

\def\PYZbs{\char`\\}
\def\PYZus{\char`\_}

\def\PYZsh{\char`\#}

\def\PYZhy{\char`\-}
\def\PYZsq{\char`\'}
\def\PYZdq{\char`\"}

\makeatother

\providecommand*{\DUrole}[2]{%
  \ifcsname DUrole#1\endcsname%
    \csname DUrole#1\endcsname{#2}%
  \else
    \ifcsname docutilsrole#1\endcsname%
      \csname docutilsrole#1\endcsname{#2}%
    \else%
      #2%
    \fi%
  \fi%
}

\ifthenelse{\isundefined{\hypersetup}}{
  \usepackage[colorlinks=true,linkcolor=blue,urlcolor=blue]{hyperref}
  \urlstyle{same} 
}{}

\begin{document}
\newcounter{footnotecounter}\title{Want Drugs? Use Python.}\author{Michał Nowotka$^{\setcounter{footnotecounter}{1}\fnsymbol{footnotecounter}\setcounter{footnotecounter}{2}\fnsymbol{footnotecounter}}$%
          \setcounter{footnotecounter}{1}\thanks{\fnsymbol{footnotecounter} %
          Corresponding author: \protect\href{mailto:mnowotka@ebi.ac.uk}{mnowotka@ebi.ac.uk}}\setcounter{footnotecounter}{2}\thanks{\fnsymbol{footnotecounter} European Molecular Biology Laboratory, European Bioinformatics Institute (EMBL-EBI), Wellcome Genome Campus, Hinxton, Cambridgeshire CB10 1SD, UK}, George Papadatos$^{\setcounter{footnotecounter}{2}\fnsymbol{footnotecounter}}$, Mark Davies$^{\setcounter{footnotecounter}{2}\fnsymbol{footnotecounter}}$, Nathan Dedman$^{\setcounter{footnotecounter}{2}\fnsymbol{footnotecounter}}$, Anne Hersey$^{\setcounter{footnotecounter}{2}\fnsymbol{footnotecounter}}$\thanks{%

          \noindent%
          Copyright\,\copyright\,2015 Michał Nowotka et al. This is an open-access article distributed under the terms of the Creative Commons Attribution License, which permits unrestricted use, distribution, and reproduction in any medium, provided the original author and source are credited. http://creativecommons.org/licenses/by/3.0/%
        }}\maketitle
          \renewcommand{\leftmark}{PROC. OF THE 8th EUR. CONF. ON PYTHON IN SCIENCE (EUROSCIPY 2015)}
          \renewcommand{\rightmark}{WANT DRUGS? USE PYTHON.}

\setcounter{page}{25}
\newcommand*{\docutilsroleref}{\ref}
\newcommand*{\docutilsrolelabel}{\label}
\AtEndDocument{\cleardoublepage}
\begin{abstract}We describe how Python can be leveraged to streamline the curation,
modelling and dissemination of drug discovery data as well as the
development of innovative, freely available tools for the related
scientific community.
We look at various examples, such as chemistry toolkits, machine-learning
applications and web frameworks and show how Python can glue it all together
to create efficient data science pipelines.\end{abstract}

\section{Introduction%
  \label{introduction}%
}

ChEMBL \cite{ChEMBL12,ChEMBL14} is a large open access database resource in
the field of computational drug discovery, chemoinformatics, medicinal
chemistry \cite{MedChem} and chemical biology.
Developed by the \href{https://www.ebi.ac.uk/chembl/}{Chemogenomics team} at the \href{http://www.ebi.ac.uk/}{European Bioinformatics
Institute}, the ChEMBL database stores curated two-dimensional chemical
structures and standardised quantitative bioactivity data alongside calculated
molecular properties.
The majority of the ChEMBL data is derived by manual extraction
and curation from the primary scientific literature, and therefore covers a
significant fraction of the publicly available chemogenomics space.

In this paper, we describe how Python is used by the ChEMBL group, in order to
process data and deliver high quality tools and services.
In particular, we cover the following topics:\newcounter{listcnt0}
\begin{list}{\arabic{listcnt0}.}
{
\usecounter{listcnt0}
\setlength{\rightmargin}{\leftmargin}
}

\item 

Distributing data
\item 

Performing core cheminformatics operations
\item 

Rapid data analysis and prototyping
\item 

Curating data\end{list}

\section{Data distribution%
  \label{data-distribution}%
}

ChEMBL offers two basic channels to share its contents:
\href{https://www.ebi.ac.uk/chembl/downloads}{SQL dump downloads} \emph{via} FTP and \href{https://www.ebi.ac.uk/chembl/ws}{web services}.
Both channels have different characteristics - data dumps are typically used by
organizations ready to host their own private instance of the database.
This method requires downloading a SQL dump file and hosting on a
machine (physical or virtual).
This approach can be expensive, both in terms of time and hardware
infrastructure costs.
An alternative approach to accessing the ChEMBL data, is to use the dedicated
web services.
This method, supported with detailed online documentation and
examples, can be used by developers, who wish to create simple widgets, web
sites, RIAs or mobile applications, that consume chemical and biological data.

The ChEMBL team uses Python to deliver the SQL dumps and web services to end
users.
In the case of the SQL dumps, the \href{https://docs.djangoproject.com/en/1.8/topics/db/queries/}{Django ORM} (Object Relational Mapping) is
employed to export data from  a production \href{http://www.oracle.com/technetwork/database/enterprise-edition/overview/index.html}{Oracle} database into two other
popular formats: \href{https://www.mysql.com/}{MySQL} and \href{http://www.postgresql.org/}{PostgreSQL}.
The \href{https://github.com/chembl/chembl_migration_model}{Django data model}, which describes the ChEMBL database schema, is
responsible for translating incompatible data types, indicating possible
problems with data during the fully automated \href{https://github.com/chembl/chembl_migrate}{migration process}.
After data is populated to separate Oracle, MySQL and PostgreSQL instances,
the SQL dumps in the respective dialects are produced.

The Django ORM is also used by the web services \cite{WS15}.
This technique simplifies the implementation of data filtering, ordering and
pagination by avoiding raw SQL statements inside the code.
The entire ChEMBL web services code base is written in Python using the
\href{https://www.djangoproject.com/}{Django framework}, \href{https://django-tastypie.readthedocs.org/en/latest/}{Tastypie} (used to expose RESTful resources) and
\href{http://gunicorn.org/}{Gunicorn} (used as an application server).
In production, Oracle is used as a database engine and \href{https://www.mongodb.org/}{MongoDB} for caching
results.
As a plus, the ORM allows for the same codebase to be used with open source
database engines.

Currently, the ChEMBL web services provide 18 distinct resource endpoints,
which offer advanced filtering and ordering of the results in JSON, JSONP,
XML and YAML formats.
The web services also support CORS, which allows them to be accessed \emph{via}
AJAX calls from web pages.
There is also an \href{https://www.ebi.ac.uk/chembl/api/data/docs}{online documentation}, that allows users to perform web
services calls from a web browser.\begin{figure*}[]\noindent\makebox[\textwidth][c]{\includegraphics[scale=0.40]{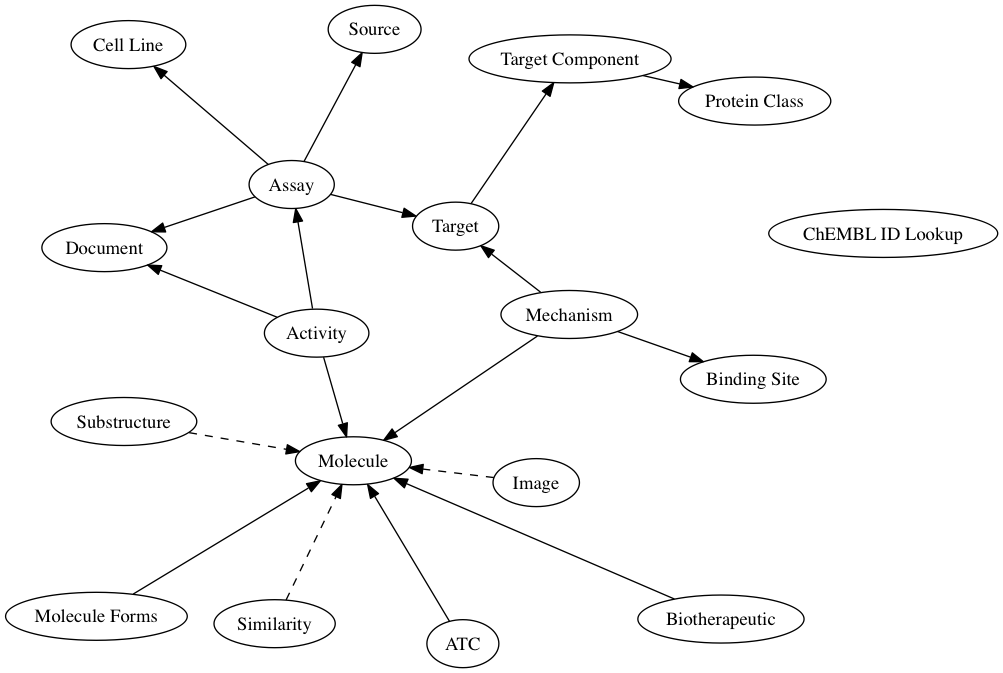}}
\caption{Diagram depicting relations between resources.
Ellipses represent ChEMBL web service endpoints and the line between two
resources indicates that they share a common attribute.
The arrow direction shows where the primary information about a resource
type can be found.
A dashed line indicates the relationship between two resources behaves
differently. \DUrole{label}{egfig}}
\end{figure*}

The \href{https://github.com/chembl/chembl_webservices_2}{web services codebase} is Apache 2.0 licensed and available from
\href{https://github.com}{GitHub}.
The code is also registered in the Python Package Index (\href{https://pypi.python.org/pypi}{PyPI}), which
allows quick deployment by third-party organizations hosting the ChEMBL
database.

\section{Performing core cheminformatics operations%
  \label{performing-core-cheminformatics-operations}%
}

There are some commonly used algorithms and methods, that are essential in the
field of cheminformatics.
These include:\setcounter{listcnt0}{0}
\begin{list}{\arabic{listcnt0}.}
{
\usecounter{listcnt0}
\setlength{\rightmargin}{\leftmargin}
}

\item 

2D/3D compound depiction.
\item 

Finding compounds similar to the given query compound with some similarity
threshold.
\item 

Finding all compounds, that have the given query compound as substructure.
\item 

Computing useful descriptors, such as molecular weight,
polar surface area, number of rotatable bonds etc.
\item 

Converting between popular chemical formats/identifiers such as SMILES,
InChI, MDL molfile.\end{list}

There are several software libraries, written in different languages, that
implement some or all of the operations described above.
Two of these toolkits offer robust and comprehensive functionality, coupled with
a permissive license, namely \href{http://www.rdkit.org/}{RDKit} (developed and maintained by Greg
Landrum) and \href{https://github.com/ggasoftware/indigo}{Indigo} (created by GGA software, now \href{http://www.epam.com/}{Epam}).
They both provide Python bindings and database cartridges, that, among other
things, allow performing substructure and similarity searches on compounds
stored in RDBMS.

The ChEMBL web services that we've described so far are focused on the retrieval
of structured data stored in databases.
Talking with colleagues, we've identified a gap in efficient pipelines, that allow
researchers to handle data process and curating chemical datasets, and we thus
focused on building additional \emph{cheminformatics-focused} services.
To fix this gap, the \href{https://github.com/chembl/chembl_beaker}{Beaker} project was setup.
Beaker \cite{Beaker14} exposes most functionality offered by RDKit using REST.
This means that the functionality RDKit provides, can now be accessed \emph{via} HTTP,
using any programming language, without requiring a local RDKit installation.

Following a similar setup to the \emph{data} part of ChEMBL web services, the \emph{utils}
part (Beaker) is written in pure Python (using \href{http://bottlepy.org/docs/dev/index.html}{Bottle framework}),
Apache 2.0 licensed, available on GitHub, registered to PyPI and has its
own \href{https://www.ebi.ac.uk/chembl/api/utils/docs}{live online documentation}.
This means, that it is possible to quickly set up a local instance of the Beaker
server.\begin{figure}[]\noindent\makebox[\columnwidth][c]{\includegraphics[scale=0.30]{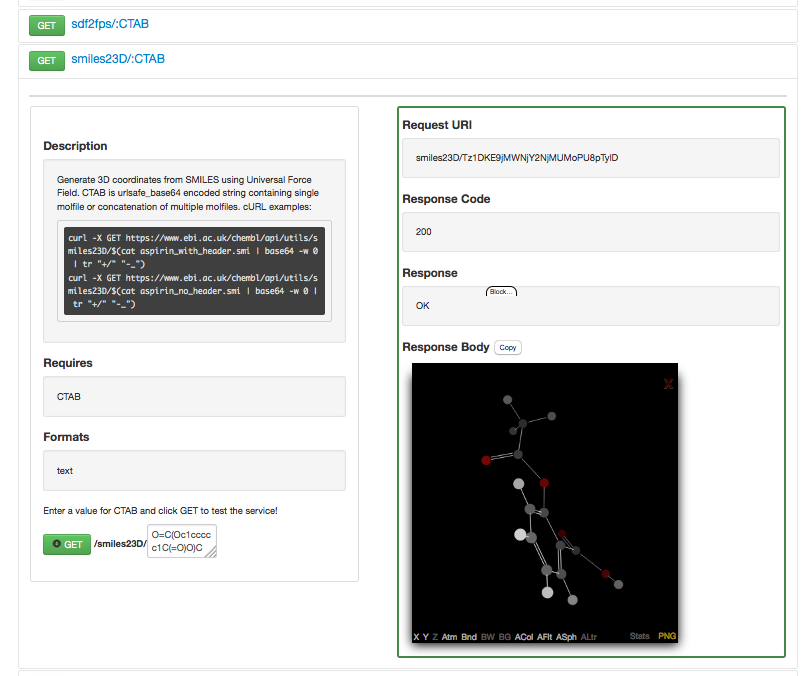}}
\caption{ChEMBL Beaker online documentation \DUrole{label}{egfig}}
\end{figure}

In order to facilitate Python software development, the \href{https://github.com/chembl/chembl_webresource_client}{ChEMBL client library}
has been created.
This small Python package wraps around \href{http://www.python-requests.org/en/latest/}{Requests library}, providing more
convenient API, similar to \href{https://docs.djangoproject.com/en/1.8/ref/models/querysets/}{Django QuerySet}, offering lazy evaluation of
results, chaining filters and caching results locally.
This effectively reduces the number of requests to the remote server, which speeds
up data retrieval process.
The package covers full ChEMBL web services functionality, allowing users
to retrieve data as well as perform chemical computations without installing
chemistry toolkits.

The following code example demonstrates how to retrieve all approved drugs for
a given target:\begin{Verbatim}[commandchars=\\\{\},fontsize=\footnotesize]
\PY{k+kn}{from} \PY{n+nn}{chembl\PYZus{}webresource\PYZus{}client.new\PYZus{}client} \PYZbs{}
    \PY{k+kn}{import} \PY{n}{new\PYZus{}client}

\PY{c+c1}{\PYZsh{} Receptor protein\PYZhy{}tyrosine kinase erbB\PYZhy{}2}
\PY{n}{chembl\PYZus{}id} \PY{o}{=} \PY{l+s+s2}{\PYZdq{}}\PY{l+s+s2}{CHEMBL1824}\PY{l+s+s2}{\PYZdq{}}

\PY{n}{activities} \PY{o}{=} \PY{n}{new\PYZus{}client}\PY{o}{.}\PY{n}{mechanism}\PYZbs{}
    \PY{o}{.}\PY{n}{filter}\PY{p}{(}\PY{n}{target\PYZus{}chembl\PYZus{}id}\PY{o}{=}\PY{n}{chembl\PYZus{}id}\PY{p}{)}
\PY{n}{compound\PYZus{}ids} \PY{o}{=} \PY{p}{[}\PY{n}{x}\PY{p}{[}\PY{l+s+s1}{\PYZsq{}}\PY{l+s+s1}{molecule\PYZus{}chembl\PYZus{}id}\PY{l+s+s1}{\PYZsq{}}\PY{p}{]}
                \PY{k}{for} \PY{n}{x} \PY{o+ow}{in} \PY{n}{activities}\PY{p}{]}
\PY{n}{approved\PYZus{}drugs} \PY{o}{=} \PY{n}{new\PYZus{}client}\PY{o}{.}\PY{n}{molecule}\PYZbs{}
    \PY{o}{.}\PY{n}{filter}\PY{p}{(}\PY{n}{molecule\PYZus{}chembl\PYZus{}id\PYZus{}\PYZus{}in}\PY{o}{=}\PY{n}{compound\PYZus{}ids}\PY{p}{)}\PYZbs{}
    \PY{o}{.}\PY{n}{filter}\PY{p}{(}\PY{n}{max\PYZus{}phase}\PY{o}{=}\PY{l+m+mi}{4}\PY{p}{)}
\end{Verbatim}
Another example will use Beaker to convert approved drugs from the previous
example to SDF file and compute maximum common substructure:\begin{Verbatim}[commandchars=\\\{\},fontsize=\footnotesize]
\PY{k+kn}{from} \PY{n+nn}{chembl\PYZus{}webresource\PYZus{}client.utils} \PY{k+kn}{import} \PY{n}{utils}

\PY{n}{smiles} \PY{o}{=} \PY{p}{[}\PY{n}{drug}\PY{p}{[}\PY{l+s+s1}{\PYZsq{}}\PY{l+s+s1}{molecule\PYZus{}structures}\PY{l+s+s1}{\PYZsq{}}\PY{p}{]}\PYZbs{}
    \PY{p}{[}\PY{l+s+s1}{\PYZsq{}}\PY{l+s+s1}{canonical\PYZus{}smiles}\PY{l+s+s1}{\PYZsq{}}\PY{p}{]} \PY{k}{for} \PY{n}{drug} \PY{o+ow}{in} \PY{n}{approved\PYZus{}drugs}\PY{p}{]}
\PY{n}{mols} \PY{o}{=} \PY{p}{[}\PY{n}{utils}\PY{o}{.}\PY{n}{smiles2ctab}\PY{p}{(}\PY{n}{smile}\PY{p}{)} \PY{k}{for} \PY{n}{smile} \PY{o+ow}{in} \PY{n}{smiles}\PY{p}{]}
\PY{n}{sdf} \PY{o}{=} \PY{l+s+s1}{\PYZsq{}}\PY{l+s+s1}{\PYZsq{}}\PY{o}{.}\PY{n}{join}\PY{p}{(}\PY{n}{mols}\PY{p}{)}
\PY{n}{result} \PY{o}{=} \PY{n}{utils}\PY{o}{.}\PY{n}{mcs}\PY{p}{(}\PY{n}{sdf}\PY{p}{)}
\end{Verbatim}

\section{Rapid data analysis and prototyping%
  \label{rapid-data-analysis-and-prototyping}%
}
Access to a very comprehensive cheminformtics toolbox, consisting of a
chemically-aware relational database, efficient data access methods
(ORM, web services, client library), specialized chemical toolkits and
many other popular general purpose, scientific and data science libraries,
facilitates sophisticated data analysis and rapid prototyping of
advanced cheminformatics applications.

This is complemented by an \href{http://ipython.org/notebook.html}{IPython notebook} server, which executes a
Python code along with rich interactive plots and markdown
formatting to improve sharing results with other scientists.

In order to demonstrate capabilities of the software environment used inside
ChEMBL a \href{https://github.com/chembl/mychembl/tree/master/ipython_notebooks}{collection of IPython notebooks} has been prepared.
They contain examples at different difficulty levels, covering following topics:\setcounter{listcnt0}{0}
\begin{list}{\arabic{listcnt0}.}
{
\usecounter{listcnt0}
\setlength{\rightmargin}{\leftmargin}
}

\item 

Retrieving data using raw SQL statements, Django ORM, web services and
the client library.
\item 

Plotting charts using \href{http://matplotlib.org/}{matplotlib} and \href{http://d3js.org/}{D3.js}.
\item 

Detailed RDKit tutorial.
\item 

Machine learning - classification and regression using \href{http://scikit-learn.org/stable/}{scikit-learn}.
\item 

Building predictive models - ligand-based target prediction tutorial using
RDKit, scikit-learn and \href{http://pandas.pydata.org/}{pandas}.
\item 

Data mining - MDS tutorial, mining patent data provided by the \href{https://www.surechembl.org/search/}{SureChEMBL}
project.
\item 

NoSQL approaches - data mining using \href{http://neo4j.com/}{Neo4j}, fast similarity search
approximation using MongoDB.\end{list}

Since many notebooks require quite complex dependencies (RDKit, numpy, scipy,
lxml etc.) in order to execute them, preparing the right environment may pose
a challenge to non-technical users.
This is the reason that the ChEMBL team has created a project called \emph{myChEMBL}
\cite{myChEMBL14}.
\href{https://github.com/chembl/mychembl/}{myChEMBL} encapsulates an environment consisting of the ChEMBL database running
on PostgreSQL engine with RDKit chemistry cartridge, web services, IPython
Notebook server hosting collection of notebooks described above,
RDKit and Indigo toolkits, data-oriented Python libraries, simple web interface
for performing substructure and similarity search by drawing a compound and many
more.\begin{figure*}[]\noindent\makebox[\textwidth][c]{\includegraphics[scale=0.30]{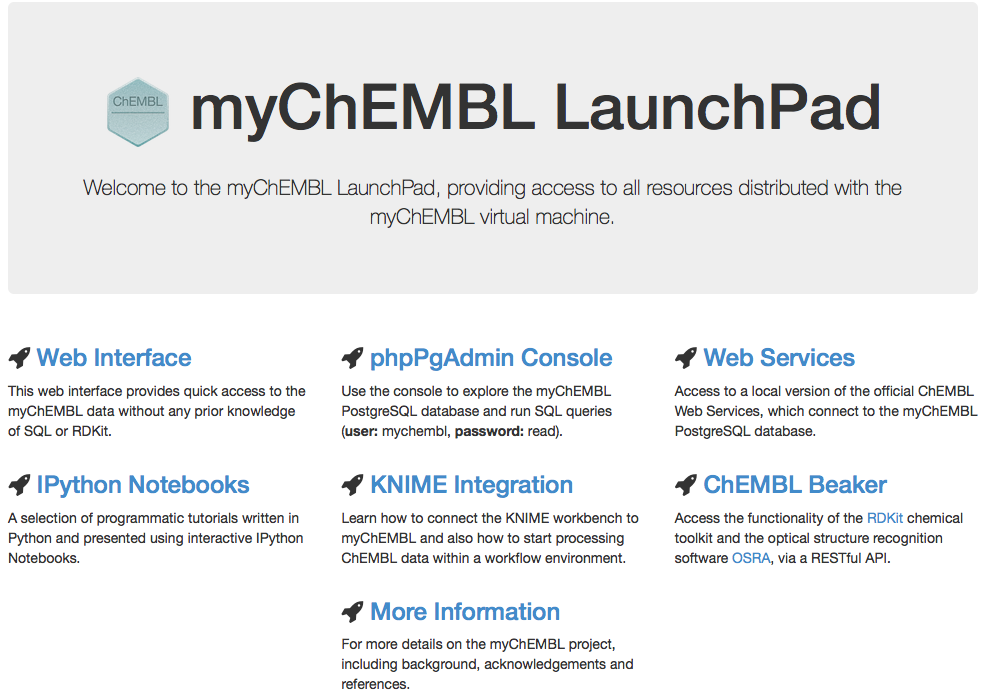}}
\caption{myChEMBL launchpad \DUrole{label}{egfig}}
\end{figure*}

myChEMBL comes preconfigured and can be used immediately.
The project is distributed as a Virtual Machine, that can be \href{ftp://ftp.ebi.ac.uk/pub/databases/chembl/VM/myChEMBL/releases/myChEMBL-20_0/}{downloaded} \emph{via}
FTP or obtained using \href{https://www.vagrantup.com/}{Vagrant} by executing the following commands:%
\begin{quote}\begin{verbatim}
vagrant init chembl/mychembl_20_ubuntu
vagrant up --provider virtualbox
\end{verbatim}

\end{quote}
There are two variants - one based on \href{http://releases.ubuntu.com/14.04/}{Ubuntu 14.04 LTS} and the second
one based on \href{https://www.centos.org/}{CentOS 7}.
Virtual Machine disk images are available in vmdk, qcow2 and img formats.
\href{https://www.docker.com/}{Docker} containers are available as well.
The scripts used to build and configure machines are available on GitHub so it
is possible to run them on physical machines instead of VMs.

Again, Python plays important role in configuring myChEMBL.
Since Docker is designed to run one process per container and ignores
OS-specific initialization daemons such as upstart, systemd etc. myChEMBL ships
with \href{http://supervisord.org/}{supervisor}, which is responsible for managing and monitoring all core
myChEMBL services (such as Postgres, Apache, IPython server) and providing a
single point of entry.

\section{Target prediction%
  \label{target-prediction}%
}

The wealth and diversity of structure-activity data freely
available in the ChEMBL database has enabled large scale data mining and
predictive modelling analyses \cite{Ligands12,Targets13}.
Such analyses typically involve the generation of classification models trained
on the structural features of compounds with known activity.
Given a new compound, the model predicts likely biological targets, based
on the enrichment of structural features against known targets in the training set.
We implemented our own classification model using:\setcounter{listcnt0}{0}
\begin{list}{\arabic{listcnt0}.}
{
\usecounter{listcnt0}
\setlength{\rightmargin}{\leftmargin}
}

\item 

a carefully selected subset of ChEMBL as a training set stored as a pandas dataframe,
\item 

structural features computed by RDKit,
\item 

the naive Bayesian classification method implemented in scikit-learn.\end{list}

As a result, ChEMBL provides predictions of likely targets for known drug
compounds available online
(e.g. in \url{https://www.ebi.ac.uk/chembl/compound/inspect/CHEMBL502}), along with the
models themselves available to download
(\url{ftp://ftp.ebi.ac.uk/pub/databases/chembl/target_predictions/}).
This is complemented with an IPython Notebook tutorial on using these models and
getting predictions for arbitrary input structures.

Furthermore, similar models have been used in a publicly available web application
called \href{https://www.ebi.ac.uk/chembl/admesarfari}{ADME SARfari} \cite{Sarfari}.
This resource allows cross-species target prediction and comparison of ADME
(Absorption, Distribution, Metabolism, and Excretion) related targets for a particular
compound or protein sequence.
The application uses \href{http://www.sqlalchemy.org/}{SQLAlchemy} as an ORM, contained within a web framework
(\href{http://www.pylonsproject.org/}{Pyramid} \& \href{https://cornice.readthedocs.org/en/latest/}{Cornice}) to provide an API and HTML5 interactive user interface.

\section{Curation of data%
  \label{curation-of-data}%
}

Supporting and automating the process of extracting and curating data from scientific
publications is another area where Python plays a pivotal role.
The ChEMBL team is currently working on a web application, that can aid in-house
expert curators with this challenging and time-consuming process.
The application can open a scientific publication in PDF format or a scanned
document and extract compounds presented as images or identifiers.
The extracted compounds are presented to the user in order to correct possible
errors and save them to database.
The system can detect compounds already existing in database and take
appropriate action.\begin{figure*}[]\noindent\makebox[\textwidth][c]{\includegraphics[scale=0.30]{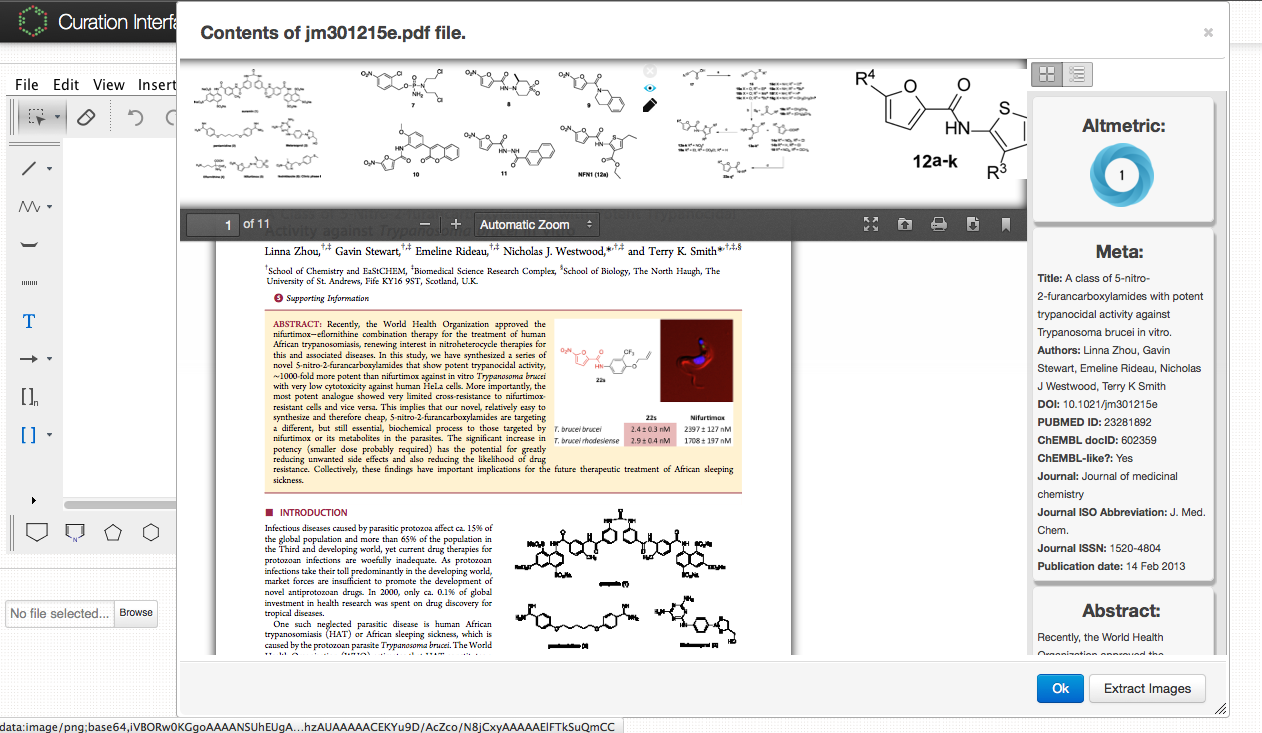}}
\caption{Extracting data from a scientific publication. \DUrole{label}{egfig}}
\end{figure*}

In addition to processing scientific papers and images, curation interface can
handle the most popular chemical formats, such as SDF files, MDL molfiles,
SMILES and InChIs.
\href{http://www.celeryproject.org/}{Celery} is used as a synchronous task queue for performing the necessary
chemistry calculations when a new compound is inserted or updated.
This system allows a chemical curator to focus on domain specific tasks and no
longer interact directly with the database, using raw SQL statements, which can
be hard to master and difficult to debug.

\section{Discussion%
  \label{discussion}%
}

Python has become an essential technology requirement of the core activities
undertaken by ChEMBL group, in order to streamline data distribution, curation
and analysis in the field of computational drug discovery.
The tools built using Python are robust, flexible and web friendly,
which makes them ideal for collaborating in a scientific environment.
As an interpreted, dynamically typed scripting language, Python is ideal for
prototyping diverse computing solutions and applications.
The combination of a plethora of powerful general purpose and scientific libraries,
that Python has at its disposal, (e.g. scikit-learn, pandas, matplotlib), along
with domain specific toolkits (e.g. RDKit), collaborative platforms
(e.g. IPython Notebooks) and web frameworks (e.g. Django), provides a complete
and versatile scientific computing ecosystem.

\section{Acknowledgments%
  \label{acknowledgments}%
}

We acknowledge the following people, projects and communities, without whom
the projects described above would not have been possible:\setcounter{listcnt0}{0}
\begin{list}{\arabic{listcnt0}.}
{
\usecounter{listcnt0}
\setlength{\rightmargin}{\leftmargin}
}

\item 

Greg Landrum and the RDKit community (\url{http://www.rdkit.org/})
\item 

Francis Atkinson, Gerard van Westen and all former and current
members of the ChEMBL group.
\item 

All ChEMBL users, in particular those who have contacted chembl-help and
suggested enhancements to the existing services\end{list}

\end{document}